\documentclass[aip]{revtex4-1}
\usepackage{graphicx}
\usepackage{amsmath,amssymb,amsfonts}
\usepackage{dcolumn}
\usepackage{bm}
\usepackage{slashed}
\usepackage{nicefrac}
\usepackage{epstopdf}
\usepackage{color}
\usepackage{tikz-feynman,contour}
\def\bra{\langle}
\def\ket{\rangle}

\def\cD{\mathcal{D}}
\def\cN{\mathcal{N}}
\def\N{\mathbb{N}}

\def\cF{\mathcal{F}}
\def\cZ{\mathcal{Z}}

\def\cH{\mathcal{H}}
\def\vx{\mathbf{x}}
\def\vk{\mathbf{k}}

\def\vb{\mathbf{b}}
\def\R{\mathbb{R}}

\def\lr{{L^2(\R)}}
\def\clos{{\rm clos\, }}
\newcommand{\Z}{\mathbb{Z}} 
\def\dk#1#2{\frac{ d^{#2}{#1} }{ (2\pi)^{#2} }} 
\newtheorem{definition}{Definition}

\begin{document}
\title{Multiresolution quantum field theory in light-front coordinates}
\author{Mikhail V. Altaisky}
\affiliation{Space Research Institute RAS, Profsoyuznaya 84/32, Moscow, 117997, Russia}
\email{altaisky@rssi.ru}
\author{Natalia E. Kaputkina}
\affiliation{National University of Science and Technology 'MISIS',  Leninsky av. 4, Moscow, 119049, Russia}
\email{kaputkina.ne@misis.ru}
\author{Robin Raj}
\affiliation{Mahatma Gandhi University,  Priyadarsini Hills, Kottayam, Kerala, 686560, India}
\email{robuka97@gmail.com}
\date{Feb 11, 2022} 
\begin{abstract}
We analyse the use of wavelet transform in quantum field theory models written in light-front coordinates. 
In a recent paper \cite{Polyzou2020} 
W.N.Polyzou used $x^+$ variable as 'time', and applied wavelet transform 
to the 'spatial' coordinates only. This makes the theory asymmetric with respect to space and time coordinates.
In present paper we generalise the concept of continuous causal path, which is the basis of path integration, 
to the sequences of causally ordered spacetime regions, and present evaluation rules for Feynman path integrals 
over such sequences in terms of wavelet transform. Both the path integrals and the wavelet transform in our 
model are symmetric with respect to the light-front variables ($x^+,x^-$). The definition of a spacetime 
event in our generalization is very much like the definition of event in probability theory.   
\end{abstract}

\maketitle

\section{Introduction \label{intro:sec}}
The Feynman path integral is the basic tool of quantum field theory (QFT) and statistical mechanics.
Having initially originated from the Dirac idea that the transition of a quantum system 
from initial state $q_i$ at the time instant $t_i$ to the final state $q_f$ at the time $t_f$ can be represented by the integral over all possible intermediate states $q_\tau,t_i<\tau <t_f,$ of 
the exponent of the Lagrangian $\exp \left(\frac{\imath}{\hbar}\int_{t_i}^{t_f}L[q(\tau)]d\tau\right)$ 
\cite{Dirac1933}. Feynman has successfully applied the Dirac method to generate a 
self-consistent Lorentz-invariant perturbation theory in quantum electrodynamics (QED) 
from an infinite-dimensional integral over all possible field configurations in Minkowski 
space \cite{Feynman1949}.

Green functions calculated by the Feynman expansion suffer from divergences and cannot be compared to experimental results, unless a special procedure, 
called {\em renormalization} is applied to get rid of UV divergences. 
A way of getting tractable results without renormalization has been proposed in a series of papers 
\cite{Alt2002G24J,Altaisky2010PRD,AK2013}. It consists in changing the functional space of local square-integrable fields 
$\phi = \phi(x)$ to the space of functions that depend on both the spacetime point ($x$) and the size of 
the region ($a$) the measurement can be made on: $\phi = \phi_a(x)$. 
For a local field an attempt 
to measure some physical quantity sharp at a point $x$, i.e., at $\Delta x\to0$, demands an infinite momentum transfer $\Delta p \sim \frac{\hbar}{\Delta x} \to \infty$, which drives the theory out of its 
applicability domain by means of UV divergences. For a scale-dependent field $\phi_a(x)$ no such problem exists 
because the observation scale $a$ always remains finite. The dependence on the observation scale is 
given by the logarithmic derivatives $a\frac{\partial}{\partial a}$ that describe the renormalization of parameters 
of the scale-dependent quantum field theory \cite{Altaisky2016PRD}. 

To define the path integral in terms 
of the fields that can be observed over a finite-size regions, we first need to define {\em what are these paths}. In standard approach, Feynman path is a causally-ordered continuous set of points of the Minkowski space, with each point ($t,\vx$) considered as a label of a potential event, such as particle creation, or  particle annihilation. This 
definition, adopted in local quantum field theory, has no respect to the {\em finiteness} of the spacetime region,
where something can be measured. To define the path integral in terms of the potentially measured quantities we need something similar to  the $\sigma$-algebra of events adopted in probability 
theory \cite{Kolmogorov1956}. At the same time the construction should be invariant under the Lorentz transformations, providing the speed of light is the same in any inertial frame of reference.

In the present paper we make an endeavour to define the causal paths in the space of events using the light-front coordinates in Minkowski space and a set of basic functions which 
provide the construction of wavelet transform in light-front variables. Our model is essentially different from usual construction of path integral in terms of the light-front variables, where the $x^+$ variable is used as a 'time', which orders the events, with the $x^-$ variables being considered as a 'spatial' coordinate \cite{KS1970}. The path integral, which 
describes the transition amplitude from the initial field configuration defined on a 
spacetime region $A$ to the final field configuration defined on a spacetime region $B$, 
so that $B$ is inside the forward light-cone of $A$, is completely symmetric with respect 
to the $x^+$ and the $x^-$ variables, and hence is symmetric with respect to space and time in any Lorentzian frame of reference. 

The remainder of this paper is organized as follows. In {\em Section \ref{sdf:sec}} we 
remind the introduction of scale-dependent functions by means of wavelet transform. 
{\em Section \ref{eucl:sec}} presents an example of a finite Euclidean quantum field theory model with 
$\phi^4$ interaction using continuous wavelet transform. In {\em Section \ref{lf:sec}} we present a construction of the generating functional of scale-dependent 
Green functions by means of discrete wavelet transform on a finite domain 
$[0,T]\otimes [0,T]$ in light-front coordinates ($x^+,x^-$). The analysis of 
the causality aspects of the presented wavelet-based construction ensures 
it is symmetric with respect to space and time variables. The path integration 
is performed over the sequences of causally connected events of finite size.  
The virtues and the problems of our approach are 
summarized in {\em Conclusion}. 

\section{Scale-dependent functions \label{sdf:sec}}
\subsection{Wavelet transform in $\lr$}
The idea of substituting a local field $\phi(x), x\!\in\!\R^d$ by a collection of local fields $\{ \phi_a(x) \}_a, a\in \R_+$ has first emerged in geophysics \cite{GGM1984}, where {\em non-local} separation of different components 
of the seismic signal by means of Fourier transform 
$$ \phi(x) \stackrel{\cF}{\to} \tilde{\phi}(k) $$
failed to provide reliable separation of localized signals with similar content of Fourier harmonics.
An alternative to Fourier transform was found in the convolution of the analysed signal $\phi(x)$ with certain 
well-localized function $\chi(x)$, usually referred to as a {\em mother wavelet}, shifted by $b$ and dilated by 
$a$:
\begin{equation}
\phi_a(b):= \int_\R \frac{1}{a} \bar{\chi}\left(\frac{x-b}{a} \right) \phi(x) dx. \label{dwt1}
\end{equation}
The functions $\phi_a(b)$ are called {\em wavelet coefficients} of the square-integrable function $\phi(x)$ with 
respect to the mother wavelet $\chi(x)$. (Following \cite{HM1998,Altaisky2010PRD} we have changed the $\lr$ normalisation $\frac{1}{\sqrt{a}}\chi\left(\frac{x-b}{a}\right)$ to the $L^1$ normalisation to make wavelet coefficients the same physical dimension as that of the original fields.) The function  $\phi(x)$ can be reconstructed from the set of its 
wavelet coefficients by means of {\em inverse wavelet transform}:
\begin{equation}
\phi(x) = \frac{1}{C_\chi} \int_{\R_+ \otimes \R} \frac{1}{a} \chi\left(\frac{x-b}{a} \right)\phi_a(b) \frac{da db}{a}.
\label{iwt1}
\end{equation}
The admissibility condition for the basic wavelet $\chi$ is rather loose: only the finiteness 
\begin{equation}
C_\chi = \int_0^\infty |\tilde{\chi}(a)|^2 \frac{da}{a} < \infty, \quad \hbox{where\ }
\tilde{\chi}(k):=\int_{-\infty}^\infty e^{\imath k x}\chi(x)dx, 
\end{equation}
is required to ensure the wavelet transform is invertible.

The convolutions \eqref{dwt1} and \eqref{iwt1} have clear physical interpretations: applying a 
'microscope' with an aperture function $\chi(x)$ we scrutinize the function $\phi$ locally 
at all possible resolutions; the convolution \eqref{iwt1} reconstructs function $\phi$ from a set of its 
components of all scales \cite{PhysRevLett.64.745}. Wavelet transform, as a separation of a function into 
a set of its scale components, can be written in both the continuous form, Eq.(\ref{dwt1},\ref{iwt1}), 
and the discrete form \cite{Daub1988}. 

\subsection{Continuous wavelet transform}
Wavelet transform is a natural generalization of the Fourier transform for the case when the {\em scaling} properties 
of the theory are important. Let $\cH$ be a Hilbert space of states of  quantum field $|\phi\ket$. 
Let $|x\ket \in \cH$ be a vector corresponding to the localization at point $x$, then 
\begin{equation}
\phi(x)=\bra x|\phi \ket \label{cr}
\end{equation}
is a coordinate representation of the field $|\phi\ket$. The Fourier transform is 
the decomposition of the field $|\phi\ket$ with respect to the representations of the translation group:
$$
\bra x|\phi\ket = \int \bra x|p\ket dp \bra p|\phi\ket,
$$
where $|p\ket$ is an eigenvector of the translation operator.

Similarly, let $G$ be a locally compact Lie group acting transitively on $\cH$, 
with $d\mu(\nu),\nu\in G$ being a left-invariant measure on $G$, then, 
any $|\phi\ket \in \cH$ can be decomposed with respect to 
a representation $\Omega(\nu)$ of $G$ in $\cH$ \cite{Carey1976,DM1976}:
\begin{equation}
|\phi\ket= \frac{1}{C_\chi}\int_G \Omega(\nu)|\chi\ket d\mu(\nu)\bra \chi|\Omega^\dagger(\nu)|\phi\ket, \label{gwl} 
\end{equation} 
where $|\chi\ket \in \cH$ is  a {\em 'mother wavelet'}, or an admissible vector, satisfying the admissibility condition 
$$
C_\chi = \frac{1}{\| \chi \|^2} \int_G |\bra \chi| \Omega(\nu)|\chi \ket |^2 
d\mu(\nu)
<\infty. 
$$
The coefficients $\bra \chi|\Omega^\dagger(\nu)|\phi\ket$ are known as 
wavelet coefficients in a general sense.

If the group $G$ is Abelian, the wavelet transform \eqref{gwl} with 
$G:x'=x+b'$ is the  Fourier transform. 
The next to the Abelian group is the group of the affine transformations 
of the Euclidean space $\R^d$:
\begin{equation}
G: x' = a R(\theta)x + b,\quad x,b \in \R^d, a \in \R_+,  \label{ag1}
\end{equation} 
where $R(\theta)$ is the $SO(d)$ rotation matrix.
Here we define the representation of the affine transform \eqref{ag1} with 
respect to the mother wavelet $\chi(x)$ as follows:
\begin{equation}
\Omega(a,b,\theta) \chi(x) = \frac{1}{a^d} \chi \left(R^{-1}(\theta)\frac{x-b}{a} \right).
\end{equation}

Thus the wavelet coefficients of the function $\phi(x) \in L^2(\R^d)$ with 
respect to the mother wavelet $\chi(x)$ in Euclidean space $\R^d$ can be written 
as 
\begin{equation}
\phi_{a,\theta}(b) = \int_{\R^d} \frac{1}{a^d} \overline{\chi \left(R^{-1}(\theta)\frac{x-b}{a} \right) }\phi(x) d^dx. \label{dwtrd}
\end{equation} 
The function $\phi(x)$ can be reconstructed from its wavelet coefficients 
\eqref{dwtrd} using the formula \eqref{gwl}:
\begin{equation}
\phi(x) = \frac{1}{C_\chi} \int \frac{1}{a^d} \chi\left(R^{-1}(\theta)\frac{x-b}{a}\right) \phi_{a\theta}(b) \frac{dad^db}{a} d\mu(\theta), \label{iwt}
\end{equation}
where $d\mu(\theta)$ is the left-invariant measure on the $SO(d)$ rotation group, usually written in terms of the 
Euler angles: $$d\mu(\theta) = 2\pi \prod_{k=1}^{d-2} \int_0^\pi \sin^k \theta_k d\theta_k.$$
The normalization 
constant
$C_\chi$ is readily evaluated using Fourier transform.
For isotropic 
wavelets
\begin{equation}
C_\chi = \int_0^\infty |\tilde \chi(ak)|^2\frac{da}{a}
= \int |\tilde \chi(k)|^2 \frac{d^dk}{S_{d}|k|^d} < \infty,
\label{adcfi}
\end{equation}
where $S_d = \frac{2 \pi^{d/2}}{\Gamma(d/2)}$ is the area of unit sphere 
in $\R^d$, with $\Gamma(x)$ being the Euler's Gamma function. Tilde means the Fourier transform: $\tilde{\chi}(k) = \int_{\R^d} e^{\imath k x} \chi(x) d^d x$.
\subsection{Discrete wavelet transform}
The convolution of continuous and differentiable mother wavelet, shifted and dilated, with the analysed function 
is not very efficient numerically. For practical analysis, thanks to the works of I.Daubechies and other authors, 
we often use the frames of orthogonal wavelets, which provides fast and efficient data processing \cite{BKR1991}.
Let us briefly remind the basics of the discrete wavelet transform \cite{Daub1988,Daub10}.

The implementation of the discrete version of wavelet transform is intimately 
related to the {\em Mallat multiresolution analysis} (MRA) \cite{Ma1986}. Having been used in signal processing for quite some time before wavelets, the multiresolution analysis  in the Hilbert space of 
square-integrable functions $\lr$,or 
\begin{definition}
The Mallat sequence, 
is an increasing
sequence of closed subspaces $\{ V_j \}_{j\in\Z}, V_j \in \lr$, such that
\begin{enumerate}
\item $ \ldots \subset V_0 \subset V_1 \subset V_2 \subset \ldots \subset L^2(\R)$
\item $\displaystyle \clos_{L^2} \cup_{j\in\Z} V_j = \lr$
\item $\displaystyle \cap_{j\in\Z} V_j = \emptyset$
\item The spaces $V_j$ and $V_{j+1}$ are "similar" in a sense that 
$$
 f(x) \in V_j \Leftrightarrow f(2x) \in V_{j+1},\quad j \in \Z.$$
\end{enumerate} 
\end{definition}
The property 4) is a key feature of the MRA sequence: If a set of functions 
$
\varphi_k^0 \equiv \varphi(x-k) 
$
forms a Riesz basis in $V_0$, it automatically implies that the set 
$$
\varphi^j_k = 2^{\frac{j}{2}}\varphi(2^{j}x-k)
$$
forms a basis in $V_j$. 
Due to the inclusion property 1), any function $f(x)\in V_0$ can be written as a sum 
of basic functions from $V_1$:
$$
f(x) = \sum_k c_k 2^{\frac{1}{2}}\varphi(2x-k).
$$
For this reason $$V_1 = V_0 \oplus W_0,$$ which is the definition of the orthogonal 
complement of $V_0$ to $V_1$. Similarly,
$$
V_2 = V_1 \oplus W_1,\quad V_2=V_0 \oplus W_0 \oplus W_1,
$$
and so on.

The basic functions $\varphi^j_k$ in $V_j$ spaces are usually referred to as the {\em scaling functions}. The basic functions in orthogonal complements $W_j := V_{j+1}\setminus V_j$ 
are referred to as {\em wavelet functions}:
$$
\chi^j_k(x) = 2^{\frac{j}{2}}\chi(2^{j}x-k).$$
The spaces $V_j$ and $W_j$ have clear physical interpretation: If $V_0$ is a space 
of functions measured with the most rough resolution $L_0$, then $V_1$ has a twice 
better resolution $L_0/2$; $W_j$ represents the details to be lost by the  
coarse-graining from  $V_{j+1}$ to $V_{j}$.

\begin{figure}[ht]
\centering \includegraphics[width=6cm]{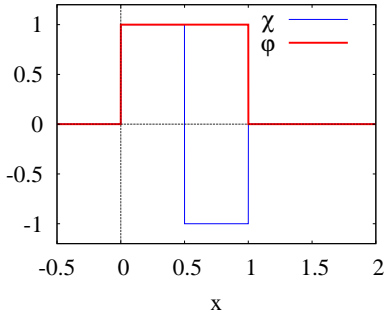}
\caption{The Haar wavelet $\chi(x)$ and its scaling function $\varphi(x)$}
\label{hp:fig}
\end{figure}

Requirements of the orthonormality of basic functions and compactness of their support 
on $[0,2N-1]$ for some $N\in \N$ enables the iterative construction of the basic wavelets from the scaling 
equation:
\begin{equation}
\varphi(x) = \sqrt{2} \sum_k h_k \varphi(2x-k),
\end{equation} 
from where the basic wavelet functions are derived \cite{Daub1988}:
\begin{equation}
\chi(x) = \sqrt{2} \sum_{k=0}^{2N-1} g_k \varphi(2x-k),\quad  g_k = (-1)^k h_{2N+1-k}. \label{hg}
\end{equation} 
The simplest wavelet with 
compact support is the Haar wavelet, which is characterized by two basic functions 
$\{ \chi^i(x)\} = \{\varphi(x),\chi(x)\}$, where  the scaling function $\varphi(x)$ \cite{Daub10}
is the indicator function of the unit interval, and $\chi(x)$ is the Haar wavelet:
\begin{equation}\varphi(x) = \begin{cases} 1 :& 0 \le x \le 1, \\
0: & \hbox{otherwise}
\end{cases},
\chi(x) = \begin{cases}
+1 ,& 0 \le x < 1/2 \cr
-1 ,& 1/2 \le x < 1 \cr 
0 ,& \hbox{otherwise}
\end{cases}. \label{haar}
\end{equation}
The graphs of the both functions are shown in Fig.~\ref{hp:fig}. 
The functions $\varphi(x)$ and $\chi(x)$, except for the simplest Haar wavelet \eqref{haar},$N=1$, defined 
by a pair of coefficients $h_0=h_1=\frac{1}{\sqrt{2}}$, usually do not have any simple 
form, but are defined recursively \cite{Daub1988}. Nevertheless, the existence of the exact 
relation \eqref{hg}, expressing the bases in the orthogonal complement spaces $W_j$ 
in terms of the scaling function allows for fast and efficient numerical algorithms, 
when each space $V_m$ is restricted to the space of functions piecewise constant on $[2^mn,2^m(n+1)]$.

Provide the chain 1) is bounded from above by the best resolution space $V_M$, consisting 
for definiteness of $2^M$ values $$(c_0^M,\ldots,c^M_{2^M-1}),$$ one can 
easily decompose this dataset into projections on 
$$W_{M-1} \oplus \ldots \oplus W_2\oplus W_1\oplus W_0 \oplus V_0$$
by applying a pair of filters ($h,g$):
\begin{equation}
c^{j-1}_i=\sum_{k=0}^{2N-1}h_k c^{j}_{k+2i},\quad 
d^{j-1}_i=\sum_{k=0}^{2N-1}g_k c^{j}_{k+2i},
\end{equation}    
where $c^j_i$ are coefficients of the projection on $V_j$,$d^j_i$ are coefficients of the 
projection on $W_j$, and periodic conditions are assumed in the discrete coordinate $i$.
 
\section{Euclidean scale-dependent QFT \label{eucl:sec}}
In ordinary quantum field theory models the field functions are defined as the scalar product 
of the field state and the localization state \eqref{cr}. If we want to define the fields that 
depend not only on the localization point $x$, but also on the scale of measurement, and possibly on 
some parameters of observation, similarly to \eqref{cr}, we can define {\em scale-dependent fields},
or scale components of the field $\phi$, as:
\begin{equation}
\phi_{a\theta}(x)\equiv \bra x,a,\theta;\chi|\phi\ket, \label{mr}
\end{equation}  
where $\bra x,a,\theta;\chi|$ is the bra vector corresponding to localization of the measuring 
device around the point $x$ with spatial resolution $a$ and the orientation $\theta \in SO(d)$.
The mother wavelet $\chi$ stands for the apparatus function of the device, or an aperture \cite{PhysRevLett.64.745}.
The scale-dependent functions \eqref{mr} can be naturally defined in terms of continuous wavelet 
transform \cite{Alt2002G24J,Altaisky2010PRD}.

To illustrate the method let us consider the theory of scalar field with $\phi^4$ interaction,
defined by the generating functional 
\begin{equation}
Z[J] = \cN \int  \cD \phi
\exp\left(-S_E[\phi] + \int J(x)\phi(x) d^dx   \right) , \label{gf1}
\end{equation} 
where 
\begin{equation}
S_E[\phi] = \int_{\R^d}\left[ \frac{1}{2} (\nabla \phi)^2 + \frac{m^2}{2}\phi^2 
+ \frac{\lambda}{4!}\phi^4
\right] d^dx \label{se4}
\end{equation}
is Euclidean action, and $\cN$ is a formal normalisation constant. The action \eqref{se4} is an isotropic, translationally-invariant functional, similar to the Ginzburg-Landau free energy functional in the theory of phase transitions \cite{GL1950}. 
It coincides with the extrapolation of the energy of a classical interacting Ising spin system to a continuous limit in $\R^d$ \cite{WK1974,BTW2002}, and describes many other physical systems. 
The connected Green functions of the theory with the Euclidean action \eqref{se4},
derived as functional derivatives of the generating functional,
\begin{equation}
G^{(n)}(x_1,\ldots,x_n) = 
\left. { \frac{\delta^n\ln Z[J]}{\delta J(x_1) \ldots \delta J(x_n)}
}\right|_{J=0},
\label{cgf} 
\end{equation}
 suffer from UV divergences. 

The straightforward way to get rid of UV divergence in Euclidean theory is to express the local field 
$\phi(x)$ in terms of the inverse wavelet transform  \eqref{iwt} and restrict the integration over all 
scale arguments to a limited range of scales $\int_A^\infty \frac{da}{a} \cdots$, considering $A$ as the 
finest resolution of observation. Unlike the momentum cutoff, the cutoff in scale argument does not produce 
any problems with the momentum conservation. 
This substitution provides a field theory model for scale-dependent fields $\phi_a(x)$ determined by the 
generating functional
\begin{equation} \label{gfw}
Z_W[J_a(x)] =\int\cD \phi_a(x) \exp\left( -S_W[\phi_a(x)] + \int \phi_a(x)J_a(x) \frac{da d^dx}{C_\chi a}\right),
\end{equation} 
where we assume $\chi$ to be an isotropic wavelet and drop the formal normalisation constant.

The action $S_W[\phi_a(x)]$ is a result of wavelet transform of all fields in the original Euclidean action 
$S_E[\phi(x)]$. For the $\phi^4$ field theory in $\R^d$ we get    
\begin{align} \nonumber 
S_W[\phi_a(x)] &=  \frac{1}{2}\int \phi_{a_1}(x_1) D(a_1,a_2,x_1-x_2) \phi_{a_2}(x_2)
\frac{da_1d^dx_1}{C_\chi a_1}\frac{da_2d^dx_2}{C_\chi a_2}  \\
&+\frac{\lambda}{4!}
\int V_{x_1,\ldots,x_4}^{a_1,\ldots,a_4} \phi_{a_1}(x_1)\cdots\phi_{a_4}(x_4)
\frac{da_1 d^dx_1}{C_\chi a_1} \frac{da_2 d^dx_2}{C_\chi a_2} \frac{da_3 d^dx_3}{C_\chi a_3} \frac{da_4 d^dx_4}{C_\chi a_4}, 
 \label{Sw}
\end{align} 
with $D(a_1,a_2,x_1-x_2)$ and $V_{x_1,\ldots,x_4}^{a_1,\ldots,a_4}$ denoting the wavelet images of the inverse propagator and that of the interaction potential \cite{Altaisky2010PRD}.

 All scale-dependent fields [$\phi_a(x)$] in 
Eq.\eqref{gfw} still interact with each other with the same coupling constant $\lambda$, but their interaction is now 
modulated  by wavelet factor $V_{x_1x_2x_3x_4}^{a_1a_2a_3a_4}$, which is the Fourier transform of $\prod_{i=1}^4 \tilde{\chi}(a_ik_i)$.

Doing so, we have the following modification of the Feynman diagram technique
\cite{Alt2002G24J}:
\begin{enumerate}\itemsep=0pt
\item Each field $\tilde\phi(k)$ is substituted by the scale component
$\tilde\phi(k)\to\tilde\phi_a(k) = \overline{\tilde \chi(ak)}\tilde\phi(k)$.
\item Each integration in the momentum variable is accompanied by the corresponding 
scale integration
\[
 \dk{k}{d} \to  \dk{k}{d} \frac{da}{a} \frac{1}{C_\chi}.
 \]
\item Each interaction vertex is substituted by its wavelet transform; 
for the $N$-th power interaction vertex, this gives multiplication 
by the factor 
$\displaystyle{\prod_{i=1}^N \tilde \chi(a_ik_i)}$.
\end{enumerate}
According to these rules, the bare Green function in wavelet representation 
takes the form
$$
G^{(2)}_0(a_1,a_2,p) = \frac{\tilde \chi(a_1p)\tilde \chi(-a_2p)}{p^2+m^2}.
$$ 
The finiteness of the loop integrals is provided by the following rule:
{\em There should be no scales $a_i$ in internal lines smaller than the minimal scale 
of all external lines} \cite{Alt2002G24J,Altaisky2010PRD}. Therefore, the integration in $a_i$ variables is performed from 
the minimal scale of all external lines up to infinity. 

The generating functional $Z_W[J_a(x)]$ is a partition function of a statistical model 
with a probability measure $e^{-S_W[\phi_a(x)]}\cD \phi_a(x)$ defining a probability of 
each field configuration $\{ \phi_a(x) {\}}_{a\in\R_+,x\in\R^d}$. The Green functions 
$$
\bra\phi_{a_1}(x_1)\cdots\phi_{a_n}(x_n)\ket_c
= \left. \frac{\delta^n \ln Z_W[J_a]}{\delta J_{a_1}(x_1)\ldots 
\delta J_{a_n}(x_n)} \right|_{J=0}, 
$$ are cumulants of the field $\phi_a(x)$.

If a given model with the 'action' $S_W[\phi_a(x)]$ was derived from a local QFT model 
with polynomial interaction, then each internal line, connecting the $i$-th and the $j$-th 
vertices of a Feynman diagram, will contain two wavelet factors 
$$
\int_A^\infty |\tilde{\chi}(a_i p)|^2 \frac{da_i}{C_\chi a_i} \times 
\int_A^\infty |\tilde{\chi}(a_j p)|^2 \frac{da_j}{C_\chi a_j},
$$
where $p$ is the momentum of the line. This results in a squared wavelet cutoff 
factors $f^2(Ap)$ in each diagram line, 
where 
\begin{equation}
f(x) = \frac{1}{C_\chi}\int_x^\infty |\tilde{\chi}(a)|^2 \frac{da}{a}
\end{equation}
for isotropic wavelets \cite{Altaisky2010PRD}. An evident normalization condition 
$f(0)=1$ corresponds to the common divergent theory in the infinite resolution limit 
$A\to0$.

This factors $f^2(Ap)$ present in each internal line suppress the UV divergences and make 
the scale-dependent QFT models with the generating functional \eqref{gfw} finite by construction. The renormalization group equations in the scale-dependent theory turn 
to be the logarithmic derivatives of the effective coupling constants, or effective 
parameters with respect to the logarithm of the observation scale 
$-\frac{\partial}{\partial \ln A}$ \cite{Altaisky2016PRD}.

As usual in functional renormalization group technique \cite{Wetterich1993}, we can introduce the effective action 
functional
\begin{equation}
\Gamma[\phi_a(x)] = -\ln Z_W[J_a(x)] + \int J_a(x) \phi_a(x) \frac{da d^dx}{C\chi a},
\end{equation}
the functional derivatives of which are the vertex functions. 

We can express it in a form of perturbation expansion:
\begin{align*}
\Gamma_{(A)}[\phi_a] &= \Gamma_{(A)}^{(0)} + \sum_{n=1}^\infty \int  
\Gamma_{(A)}^{(n)}(a_1,b_1,\ldots,a_n,b_n)
\phi_{a_1}(b_1)
\ldots \phi_{a_n}(b_n) \frac{da_1d^db_1}{C_\chi a_1}
\ldots \frac{da_nd^db_n}{C_\chi a_n}
\end{align*}
The subscript $(A)$ indicates the presence in the theory of 
minimal scale -- the observation scale.

In analytical calculations it is convenient to use derivatives of the Gaussian as 
mother wavelets.  The simplest is the first derivative of the Gaussian, which has  Fourier 
transform 
\begin{equation}
\tilde{\chi}_1(k) = -\imath k e^{-\frac{k^2}{2}}.
\end{equation}
This results in the wavelet cutoff factor $f_{\chi_1}(x)=e^{-x^2}$.
Considering the $\phi^4$ model \eqref{Sw} in one-loop approximation 
\begin{equation}
\Gamma^{(4)} = -
\begin{tikzpicture}[baseline=(a)]
\begin{feynman}[inline=(a)]
    \vertex [dot] (a);
    \vertex [above=1cm of a] (i1){\(1\)};
    \vertex [left =1cm of a] (i2){\(2\)};
    \vertex [right=1cm of a] (i3){\(3\)};
    \vertex [below=1cm of a] (i4){\(4\)};
    \diagram*{  
    {(i1),(i2),(i3),(i4)} -- [scalar] (a),
    };
\end{feynman}      
\end{tikzpicture}
  -\frac{3}{2}
  \begin{tikzpicture}[baseline=(e)]
  \begin{feynman}[horizontal = (e) to (f)]
  \vertex [dot] (e);
  \vertex [dot,right=1.5cm of e] (f);
  \vertex [above left = 1cm of e] (i1) {\(1\)};
  \vertex [below left = 1cm of e] (i2) {\(2\)};
  \vertex [above right = 1cm of f] (i3) {\(3\)};
  \vertex [below right = 1cm of f] (i4) {\(4\)};
  \diagram*{
  {(i1),(i2)} -- [scalar] (e),
  {(i3),(i4)} -- [scalar] (f),
  (e) -- [scalar,half right, momentum=\(q\)] (f),
  (e) -- [scalar,half left] (f),
    };
  \end{feynman} 
  \end{tikzpicture}  
  \label{G4:fde}
\end{equation}
In four dimensions in the relativistic limit $s^2 \gg 4m^2$ we get the following scaling equation for the coupling constant $\lambda = \lambda^{eff}(A)$:
\begin{equation}
\frac{\partial \lambda}{\partial\mu} = \frac{3\lambda^2}{16\pi^2} \frac{1-e^{-\alpha^2}}{\alpha^2}
e^{-\alpha^2}, \label{b1}
\end{equation}
where $\mu = -\ln A + const, \quad \alpha = As,\quad s = p_1+p_2$. The details of calculations are given in \cite{Altaisky2016PRD,Altaisky2010PRD}.

In the limit of infinite resolution ($A\to0$) the scaling behaviour of the coupling 
constant \eqref{b1} coincides with standard RG result. As it was shown in the recent 
paper \cite{AR2020}, the same is true for quantum electrodynamics: the asymptotic behaviour at small scales 
is independent on the particular type of the mother wavelet $\chi$, and coincides with 
the known RG behaviour. For finite scales $A>0$, the type of the mother wavelet $\chi$ 
certainly matters, because the scale components $\phi_a(x)$ have been defined with 
respect to a given $\chi$.   

\section{Scale-dependent QFT in light-front coordinates \label{lf:sec}}
\subsection{Geometric issues}
Euclidean QFT models supplied with wavelet decomposition of the fields into scale components
$\phi_a(x)$, as described in the previous section, provide finite Green functions 
for the fields defined on finite regions in $\R^d$, but these regions cannot be directly 
interpreted as spacetime regions. In the Minkowski space, in contrast, we cannot define 
a vicinity of a spacetime point $x\in \R^{1,d-1}$ using a single mother wavelet. 
This is because the group $SO(1,1)$ of  Lorentz  transformations  of  pseudo-Euclidean plane 
$\R^{1,1}$ is not a simply-connected group, but includes four connected components 
\begin{align*}
\begin{pmatrix}
\cosh \eta & \sinh \eta \cr 
\sinh \eta & \cosh \eta
\end{pmatrix}&,& 
\begin{pmatrix}
\cosh \eta &-\sinh \eta \cr 
\sinh \eta &-\cosh \eta
\end{pmatrix}&,& 
\begin{pmatrix}
-\cosh\eta & \sinh \eta \cr 
-\sinh \eta &\cosh \eta
\end{pmatrix}&,&
\begin{pmatrix}
-\cosh \eta & -\sinh \eta \cr 
-\sinh \eta & -\cosh \eta
\end{pmatrix}, 
\end{align*} 
parametrized by the Lorentz boost angle $\tanh(\eta)=v/c$ -- the {\em rapidity}. The light cone boundaries 
between these domains are unpassable for  Lorentz boosts. Thus, according to 
\cite{PG2012}, instead of a single mother wavelet $\chi(x)$, the wavelet transform 
in pseudo-Euclidean plane requires {\em four} separate mother wavelets 
\begin{equation}
\chi_j(x) = \int_{A_j} \frac{d\omega dk}{(2\pi)^2} \tilde{\chi}(k)e^{-\imath (\omega t-kx)}, \label{chij}
\end{equation}
different from each other by their support in momentum space:
\begin{align} \nonumber
A_1: |\omega|>|k|, \omega>0  &,& A_2 : |\omega|>|k|, \omega <0, \\
A_3: |\omega|<|k|, \omega>0  &,& A_4 : |\omega|<|k|, \omega <0. \label{pg11}
\end{align}

Possible solution for tracing the propagation of a quantum particle from a finite spacetime region of its preparation to the finite region of its registration, in a way 
compatible with  Lorentz invariance, is the use of light-front coordinates. {\em To some extent, what we want is a Lorentz-invariant theory 
with the spacetime regions being spanned by some wavelet basis in a way totally symmetric 
with respect to the space and the time variables.}

\subsection{Causality issues}
\begin{figure}[ht]
\centering \includegraphics[width=4cm]{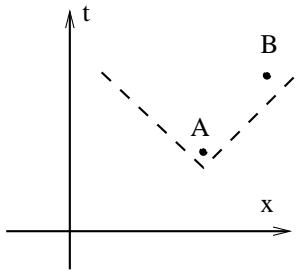}
\caption{Event $A$ can causally affect event $B$ only within the future-directed light cone.}
\label{lca:pic}
\end{figure}

We need a method  to construct path integration over all causal paths from the finite-size region 
the field was prepared to the finite-size region of its registration. 
In standard quantum field theory approach, where the Feynman path of integration is understood 
as a continuous set of spacetime points, there is only one type of causality -- the signal 
causality. Two events happening at two distinct spacetime points $x$ and $x'$, 
separated by a space-like interval $(x-x')^2<0$ cannot be causally connected.
If two events $A$ and $B$ are separated by a time-like interval $(x_A-x_B)^2>0$, 
then the event $A$ causally affects the event $B$ if $B$ is within the forward light cone 
of $A$, see Fig.~\ref{lca:pic}.
In reality, if we understand {\em an event} as a change of the matter fields that {\em can be measured} (at least in principle); the events should be defined on finite-size regions, rather than points. 
This means, that beyond the signal causality, when two events $A$ and $B$ have an 
empty intersection $A\cap B=\emptyset$, there is another option: $A\subset B$, i.e, 
$A$ is a part of $B$.
Surprisingly, we are not very certain about what is the vicinity of an event happening at 
a {\em point} $x$ in the Minkowski space \cite{Blokhintsev1973}. On one hand, it should be limited by the length of 
signal propagation during the event $\Delta x \sim c \Delta t$. On the other hand, it 
should be limited by quantum uncertainty of the event measurement $\Delta p \Delta x \sim \frac{\hbar}{2}$.

  Since each of the events is associated with some measuring process, 
the latter case is easy to imagine in a {\em non-relativistic settings}. Let $B$ be some measurement on a nucleon, and let $A$ be some measurement on a constituent quark of that nucleon. 
The measurement on a quark (which is a part of the nucleon) is enabled by means of some 
preparation, or a measurement procedure, which have been performed on the nucleon.
For instance, if the measured spin projection of nucleon to a given axis is $+\frac{3}{2}$, 
it is completely impossible for either of its constituents quarks  have to have the 
projection of spin $-\frac{1}{2}$ to the same axis. Thus, the knowledge of the state of the 
whole constrains possible states of its parts, i.e., the state of the whole is a {\em cause} for 
the states of its parts. Exactly this type of causality is manifested in the EPR-type experiments \cite{EPR1935}, when measuring the spin of one particle we automatically get the value 
of its space-like separated remote partner's spin -- tacitly exploiting the fact these two have been the parts of 
a common parent particle of spin zero.

It is rather easy to imagine such measurements in a non-relativistic case: the measurement 
domain of the part is just {\em inside} the measurement domain of the whole and the time difference between the measurements is regarded to be negligible. However, it is not 
easy to describe this situation in relativistic case. The reason is that the measurements 
performed at the same time for one observer, will be regarded as happening at different 
times for another observer. This also implies the problem of defining the  {\em event vicinity} in Minkowski space. In Euclidean space a point $x \in \R^d$ has a $\epsilon$-vicinity given by the Euclidian metrics: 
$$
V_\epsilon(x) = \{ y\in \R^d | \sqrt{(x-y)^2} \le \epsilon \}.$$ 
In Minkowski space, if we use the interval 
$$s_{xy} = \sqrt{g_{\mu\nu}(x^\mu-y^\mu)(x^\nu-y^\nu)},\quad g_{\mu\nu}=\mathrm{diag}(1,-\mathbf{1}),$$ 
as a distance between $x$ and $y$, it will be positively defined only within the future 
light-cone $V^+(x)$ -- the set of points which can be causally affected by $x$, and the past light-cone  $V^-(x)$ -- the set of points that can causally affect $x$. Thus we can 
separately define the future $\epsilon$-vicinity and the past $\epsilon$-vicinity of $x$:
\begin{align}
V^+_\epsilon(x) = \{ y \in \R^{1,3} | y^0-x^0 \ge 0, 0 \le (x-y)^2 \le \epsilon^2 \}, \\
\nonumber 
V^-_\epsilon(x) = \{ y \in \R^{1,3} | y^0-x^0 \le 0, 0 \le (x-y)^2 \le \epsilon^2 \},
\end{align}
with the only intersection $V^+_\epsilon(x) \cap V^-_\epsilon(x) = x$.
The points belonging to the light cone are separated from $x$ by zero (light-like) interval 
$(x-y)^2=0$. In this sense, any event on the light cone is indistinguishable from $x$, since 
in the coordinate system moving at the speed of light it is represented by the same point.

\begin{figure}[ht]
\centering \includegraphics[width=6cm]{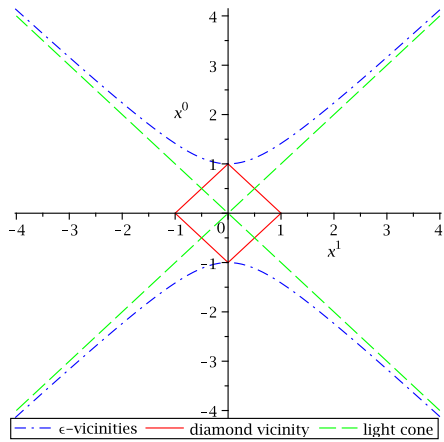}
\caption{Forward and backward vicinities of the event $P=(0,0)$ drawn in arbitrary units 
$\epsilon = \frac{\Delta t}{2} =1$. $V^\pm_\epsilon(P)$ are shown in dash-dot line. Dashed lines indicate the light cone $x^1 = \pm x^0$.}
\label{lcv:pic}
\end{figure}

The vicinity $V_\epsilon(x) := V^+_\epsilon(x) \cup V^-_\epsilon(x)$ is inconvenient for 
description of intersecting of events, one of which is the subset of the other. If we observe an event at $P=(x^0,x^1)$ during the time interval $\Delta t$ in a rest frame, our observation 
starts at $P_-=(x^0-\frac{\Delta t}{2},x^1)$ and ends at $P_+=(x^0+\frac{\Delta t}{2},x^1)$. 
This means the beginning of our observation potentially affects the forward light cone of 
$P_-$, thus the space-like separated events at the middle time $x^0$ from $(x^0,x^1-\frac{c\Delta t}{2})$ to $(x^0,x^1+\frac{c\Delta t}{2})$ may be correlated. This suggests to chose 
the diamond $D_{(x^0-\frac{\Delta t}{2},x^1)}^{(x^0+\frac{\Delta t}{2},x^1)}$ -- that is the 
intersection of causal future of $(x^0-\frac{\Delta t}{2},x^1)$ with the causal past of 
$(x^0+\frac{\Delta t}{2},x^1)$ -- as a causal $\Delta t$-vicinity of the event $P=(x^0,x^1)$. This set of vicinities is symbolically shown 
in Fig.~\ref{lcv:pic}.

In  general relativity theory a {\em causal path} connecting two spacetime events is a 
continuous curve $x^\mu = x^\mu(s)$, whose tangent vector is time-like or null everywhere 
$
\frac{dx^\mu}{ds}\frac{dx_\mu}{ds}\ge0 
$ \cite{Witten2020}. That is a causal path is a continuous set of spacetime points such that $x^\mu(s) \prec x^\mu(s')$ if $s<s'$. 

The concept of causal paths has been already extended in mathematics to  discrete 
sequences of non-empty regions \cite{CC2005,HFRW2020}: 
\begin{equation}
A_1 \prec A_2 \prec \ldots \prec A_n,
\end{equation}
where $A_i \prec A_j$ means that $\forall x\in A_i, y\in A_j, x \prec y$, in the sense that 
$y$ is in the forward light cone of $x$. The region causality relations \cite{CC2005} are based on two binary partial order relations: the preceding $\prec$ and the subset $\subset$ relations.

\begin{definition}
A set of regions $A,B,C,D,\ldots \in \cZ$ with two partial orders, such that:
\begin{enumerate}
\item The subset relation $\subset$ is a partial order on the set of regions:\\
  $A\subset B$ and $B\subset C$ implies $A\subset C$,\\
 $A\subset A$,\\
  $A\subset B$ and $B\subset A$ implies $A=B$
\item The partial order $\subset$ has a minimum element $\emptyset$, which is contained in any region.
\item The partial order $\subset$ has unions:\\
$A \subset A \cup B$ and $B \subset A \cup B$, \\
if $A\subset C$ and $B\subset C$ then $A\cup B \subset C$
\item The preceding relation $\prec$ induces a strict partial order on the non-empty regions:\\
$A\prec B$ and $B \prec C$ implies $A\prec C$, \\
$A \nprec A.$ 
\item $\forall A,B,C$:\\
  $A\subset B$ and $B\prec C$ implies $A\prec C$,\\
  $A\subset B$ and $C\prec B$ implies $C\prec A$,\\
  $A \prec C$ and $B \prec C$ implies $A\cup B \prec C$.
\end{enumerate}
is called a causal site.
\end{definition} 
(Here we have simplified the definition 2.2 given in \cite{CC2005}, reducing some requirements, redundant for the present consideration.) Causal sites generalize the concepts of 
topology using the background of the {\em category theory} \cite{MM1992}. They were intended to 
describe quantum geometry in the settings of quantum gravity, but they were not yet related 
to any measurement procedure aimed for the quantum fields defined on regions.

\begin{figure}[ht]
\centering \includegraphics[width=5cm]{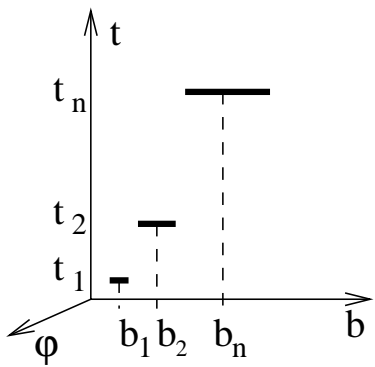}
\caption{Supports of the basic functions with time $t$ treated as an independent 
parameter. Each event also extends in the direction of scale $a$, but this is not shown. The typical spatial widths of each measurements are $a_1,a_2,a_3$, respectively. The time 
duration of each measurement is assumed to be negligible.}
\label{ht2:pic}
\end{figure}

To adopt ''the whole -- the part'' causality relations for an abstract measurement procedure and to define 
the integration over causal paths let us start from the consideration of non-relativistic case. 
In non-relativistic case, when the coordinate $x$ takes its values in Euclidean space 
$x\in\R^d$, and the time $t$ is a universal parameter, there exists a direct analogy of 
the scale-dependent fields $\phi^i_a(b)$. This is the Kadanoff blocking of the Ising 
ferromagnet \cite{Kadanoff1966}. The test function is then the indicator function of a 
unit $d$-dimensional cube. 
The $n$-point correlation functions 
$$
\bra \phi^{(i_1)}_{a_1}(b_1;t_1) \ldots \phi^{(i_n)}_{a_n}(b_n;t_n) \ket  
$$
are understood as the expectation of joint measurement of the spins, or the magnetic moments, 
of the blocks of different sizes $a_1,\ldots,a_n$, centred at $b_1,\ldots,b_n$, 
measured at different time instants $t_1,\ldots,t_n$. 
The corresponding set of such events is graphically shown in Fig.~\ref{ht2:pic}.
At coinciding time arguments the correlators can be evaluated using the ensemble averaging 
common in statistical physics. 
The set of events, corresponding to the measurements of the field values $\phi_{a_i}(b_i)$, has 
no mutual intersection of events. 

\begin{figure}[ht]
\centering \includegraphics[width=4cm]{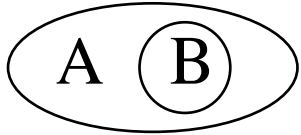}
\caption{Events in probability theory}
\label{incl:pic}
\end{figure}

In other settings, when the measured fields do not depend on time -- this happens in ergodic 
conditions when time averaging can be substituted by ensemble averaging, and only the 
spatial dependence is essential -- there may be another relation between two measurements. 
The latter is shown in Fig.~\ref{incl:pic}. 
For instance, let the event $A$ consist in measuring the fluctuations of certain field $\phi$ over a sufficiently large spatial domain (let it be, for definiteness, the turbulent fluid 
velocity as described in \cite{AHK2018}). Let the event $B$ consist in measuring the  
velocity fluctuations in some smaller domain $B$, which is inside the domain $A$. The typical scales 
of fluctuations $a$ in the smaller domain are also within the range of fluctuations in the  
larger one. In this case, the probability of a given field configuration in a smaller domain 
obeys the chain rule:
\begin{equation}
P(\phi_B) = \int P(\phi_B|\phi_A) P(\phi_A) \cD\phi_A.
\end{equation}
Thus, our measurement events turn to be the {\em events} in the common sense of an  
event algebra in probability theory \cite{Kolmogorov1956}. 

Any real physical measurement takes place in a certain spacetime domain of  finite size. 
The idealization  shown in Fig.~\ref{incl:pic}, related to {\em space} only, obtained by an ensemble averaging, is impossible 
for instrumental measurement, because the limit $\Delta t \to 0$ cannot be achieved. Hence, we face a topological question: What sequences 
of spacetime regions can comprise a path in the space of measurements, compatible 
with both the relativistic invariance an the event algebra of the probability theory?

\begin{figure}[ht]
\centering \includegraphics[width=5cm]{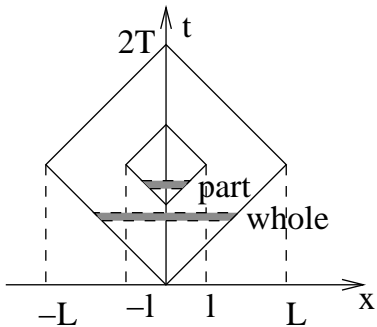}
\caption{Sequential measurements of an electron inside an atom performed in the rest frame of the atom}
\label{lcm:pic}
\end{figure}

The Fig.~\ref{incl:pic}, which totally ignores the finite time of measurement, but calculates 
the probability of the measurement result on a smaller region ($B$) {\em conditionally} to 
the state of fields on its parent region ($A$), is too strong exaggeration for the quantum 
world. Each quantum measurement is associated with a certain physical interaction. The lower is the energy of this interaction, the higher is the uncertainty in the coordinate and the 
longer is time of measurement. To measure the state of an electron in atom, we first localize the 
atom by certain {\em soft measurement} of typical size $L$, and only then perform the measurement 
on the electron itself. The time and the space uncertainty in the electron position may be kept 
deeply within the atom position $l \ll L$, see Fig.~\ref{lcm:pic}.

Even if the atom localization starts with a point interaction at $(0,0)$ and continues 
for the time $2T$, the measurement  can be affected by the 
information from the whole spacetime diamond of size $2T$ shown in  Fig.~\ref{lcm:pic}.
Thus the shape of any spacetime region that pretends to be a measurement event is 
restricted by the causal diamond of the measuring setup.  The distance 
between the beginning of the atom preparation and the measurement performed on the electron 
is given in the rest frame of the atom. However, the time difference between the different events depends 
on the choice of Lorentz frame. 

There are at least two ways to describe such hierarchic measurements. First, if the initial 
configuration of the measured system have been prepared of some equal-time hypersurface 
$\Sigma_{t_0}$ at the time instant $t_0$, we can slice the Minkowski spacetime 
into an ordered  sequence of hypersurfaces 
$\Sigma_{t_0} \prec \Sigma_{t_1} \prec \ldots \Sigma_{t_n}$, such that $t_0 < t_1 < \ldots < t_n$. This instant-type description is based on a partially-ordered set: 
$
 B \subset A \subset \Sigma_t $ implies that $A$ is a cause of $B$,   
but if the sets $A \cap B = \emptyset$ are separated by a space-like interval they cannot be 
causally ordered. 
To build a {\em continuous} trajectory of such sets $A_1 \prec A_2 \prec \ldots$ the 
distance between sequential sets ought to vanish:
\begin{equation}
A_1 \prec A_2,\lim_{t_2-t_1\to 0}d(A_1,A_2) = 0, \label{polo} \quad \hbox{where\ }
d(A_1,A_2) := \sqrt{\min |(t_1-t_2)^2-(x_1-x_2)^2|}. 
\end{equation} 
The quantization scheme based on the ordering \eqref{polo} is quite straightforward: 
the time-slices $\Sigma_t$ are ordered according to the time argument; the field operators 
corresponding to different domains of the same slice are ordered according to the rule 
''the bigger domain acts on vacuum first'' \cite{Altaisky2010PRD}.
However since the size of the time lag $\Delta t = t_n - t_{n-1}$ has no relations to 
the spacial size of the domains $A_i$ this procedure does not manifest Lorentz covariance 
\cite{Polyzou2020}. 
 
Second, to construct a Lorentz-invariant description, i.e., to be able to calculate the quantum transition amplitudes independently of the Lorentz frame, we can change the coordinates to the light-front form ($x^+,x^-,x^2,x^3$):
\begin{equation}
\begin{pmatrix}
x^+ \cr x^-
\end{pmatrix} = \frac{1}{\sqrt{2}} \begin{pmatrix}
1 & 1 \cr 1 & -1
\end{pmatrix} \begin{pmatrix}
x^0 \cr x^1
\end{pmatrix}. \label{KS}
\end{equation}
This implies a metrics 
$$
g_{\alpha\beta} = \begin{pmatrix}
0 & 1 & 0 & 0 \cr
1 & 0 & 0 & 0 \cr
0 & 0 &-1 & 0 \cr
0 & 0 & 0 &-1
\end{pmatrix},
$$
so that $x_+=x^-,x_-=x^+, x_{2,3}=-x^{2,3}$,
$
x^2 = (x^0)^2 - (x^1)^2 - (x^2)^2 - (x^3)^2 = 2 x^+ x^- -(x^\perp)^2.
$
The factor $\sqrt{2}$ in the definition \eqref{KS} corresponds to the Kogut and Soper convention 
\cite{KS1970}, and can be lifted with appropriate rescaling of $g_{\alpha\beta}$. 

To make our causal description totally symmetric with respect to the space ($x^1\equiv x$) 
and the time ($x^0\equiv t$) coordinates, we can consider the following construction. 
Let the observer be capable of observing a region of typical time span $2T$, and hence the maximal spatial size $2cT$, and let his best resolution be $\Delta t=T$, i.e., the observer 
is capable of discriminating the observed object to be in the left, or right part of the observed domain -- in the space or in the time direction, respectively.
\begin{figure}[ht]
\centering \includegraphics[width=8cm]{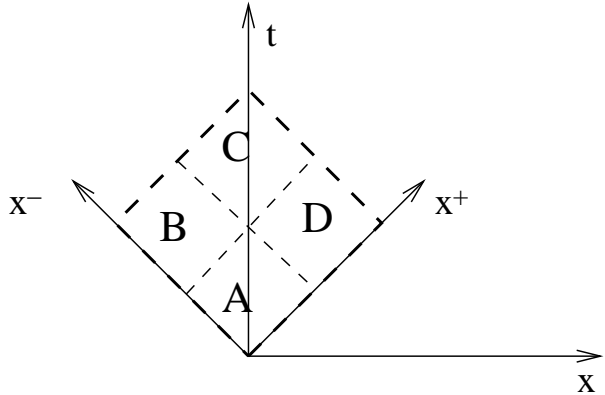}
\caption{Binary discriminated region  $\Theta$ in Minkowski plane. If the system was initially prepared in the domain $A$ and finally registered in the domain $C$, there are 
3 possible trajectories between these regions: $(ABC),(ADC),(AC)$. The light-front coordinates $x^+ = \frac{t+x}{\sqrt{2}},x^-=\frac{t-x}{\sqrt{2}}$ enable symmetric 
partition of the spacetime region $\Theta$.}
\label{xpm2:pic}
\end{figure}
The standard (signal) ordering implies a partial order on the set 
$\Theta=A\cup B \cup C \cup D$, see Fig.~\ref{xpm2:pic}:
$$
A \prec B \prec C, A \prec D \prec C.
$$ 
The regions, $B$ and $D$, being separated by a space-like interval, are not causally ordered. In the picture shown in Fig.~\ref{xpm2:pic}, the $B$ and $D$ are simultaneous, but in other Lorentz frames it may be either $B\prec D$, or $D\prec B$. 

To define paths on the set of regions independently of the Lorentz frame we need a 
set of test functions, which 
characterise each event region. It is convenient to define such basic 
functions in terms of the light-front variables. 
Let us consider a wavepacket of constant shape $\phi(\cdot)$ moving in the positive 
direction of the $x$ axis at the speed of light ($c=1$): $\phi = \phi(t-x)$. 
This wavepacket can scan the spacetime points in the vicinity of the light-front 
$t\approx x$, and the only way to discriminate the records of such events is to 
use the complementary variable 
$x^+ =\frac{t+x}{\sqrt{2}}$ as a {\em time coordinate}. Launching similar packets 
from different ''space locations'' $x^-$ one can scan different spacetime regions.
Perhaps, by voluntary choice of the right-moving direction $x^-$, we have violated 
the symmetry between the left-moving and right-moving waves, but in this settings we can use the complementary variable $x^+$ 
as the time coordinate for the construction of the Feynman functional integration. 
This change of variables is advantageous in collider experiments, with their preferable direction of the beam of relativistic particles \cite{Chen2019,Berges2021}. 

In the light-front variables $(x^-,x^+)$ the diamond, shown in Fig.~\ref{xpm2:pic}, turns 
to be a rectangle, the points of which can be ordered along the ''time'' direction $x^+$.
Having this done, we can construct the equal-$x^+$ commutation relations for the fields 
at different $x^-$-locations. 

\begin{figure}[ht]
\centering \includegraphics[width=4cm]{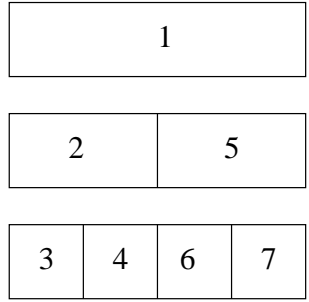}
\caption{Ordering of hierarchical events on a one-dimensional interval.}
\label{t4:pic}
\end{figure}

In the case of the scale-dependent theory, when the causal trajectory is understood as a 
sequence of possible measurements performed on different spacetime regions, the operators defined on spacetime 
regions should first be ordered with respect to the subset relation ($\subset$), and only 
then according to the signal causality relation ($\prec$). For instance, if we observe a quantum system described by a single coordinate $q=q(t)$ during the time interval $t\in [0,T)$ with the best time resolution $\Delta t = \frac{T}{4}$, the operators should 
be ordered as shown in Fig.~\ref{t4:pic}. 
First, we locate the system, or a particle, at the beginning of the trajectory by registering its existence 
on intervals $1\to2\to3$, then the movement $3\to4$ is registered. The transport of the system to the next time cell '6' is registered in two steps: first the particle is located in the right half-interval '5', and then located in the left quarter '6'. The physical 
fields characterising such system $q_{[0,T)},q_{[0,T/2)},q_{[T/2,T)},\ldots$ can be measured using the waves of different frequencies. 
\subsection{Path integration over the event regions}
To illustrate the path integration over the fields that depend on regions, rather 
than points  let us consider 
a simple quantum system with a single degree of freedom, say a harmonic oscillator. 
In quantum mechanics 
the amplitude of a quantum transition of such system from initial quantum state $|Q\ket$ at 
time $t=0$ to the final quantum state $|Q'\ket$ at time $t=T>0$ is given by 
the equation 
\begin{equation}
\bra Q'|Q\ket \propto \int \cD q e^{\frac{\imath}{\hbar} \int_0^T L[q(t)]dt}, 
\label{fi}
\end{equation}
where $L[q]$ is the Lagrangian of the system, and $\cD q$ is the Feynman's measure. 

To calculate the transition amplitude \eqref{fi} the time interval $[0,T]$ should be divided 
into small time-slices $0<t_1<\ldots t_n < T$, so that in infinitesimal limit 
($n\!\to\!\infty$)
$$
\cD q = \prod_{i=1}^n dq(t_i), \quad \max_i (t_i-t_{i-1})\to 0.
$$

Since $q(t)$ can be considered as a quantum field on the one-dimensional interval $[0,T]$, 
we can introduce the scale-dependent fields, defined on finite-size regions, using wavelet transform. 
For the compact manifold $[0,T]$ it is natural to use the {\em discrete wavelet transform} 
using compactly supported wavelets \cite{Daub1988,Daub10}. 
In case of the discrete wavelet transform \cite{Daub1988} the integrals  over the scale variable $\frac{da}{a}$ 
become discrete sums $\sum_m d^m_n \chi_{mn}(t)$, where  
\begin{equation}
\chi_{mn}(t):= a_0^{-\frac{m}{2}} \chi \left(
a_0^{-m}t - n b_0
\right),
\end{equation}
with $a_0=2$ being the common choice. 
The discrete set of wavelet coefficients  is given by 
\begin{equation}
d^m_{n} := \int_0^T a_0^{-\frac{m}{2}} \bar\chi \left(
a_0^{-m}t - n b_0
\right) q(t) dt \equiv \bra \chi_{mn}|q\ket. \label{dwt}
\end{equation}  

For simplicity let us consider the field $q(t)$ on the time interval $[0,T]$ written 
as a sequence of $2^N$ values  $q_0,q_1,\ldots,q_{2^N-1}$, at sequential time instants, where $q_0=q(0)$ and 
 $q_{2^N-1}=q(T)$. Using the discrete wavelet transform, with the simplest orthogonal wavelet -- the Haar wavelet \eqref{haar} -- we have the mapping 
 $$
 q_0,\ldots,q_{2^N-1} \to d^1_0\ldots d^1_{2^{N-1}-1}, d^2_0\ldots d^2_{2^{N-2}-1},
 \ldots d_0^N c_0^N,
 $$
given by the iterative procedure of the fast wavelet transform algorithm
\begin{equation}
c_k^{j+1} = \frac{c^j_{2k}+c^j_{2k+1}}{\sqrt{2}}, \quad 
d_k^{j+1} = \frac{c^j_{2k}-c^j_{2k+1}}{\sqrt{2}},\quad j\in \N \label{fwt}
\end{equation}
with the initial data $q_k \equiv c^0_k$. 
The algorithm \eqref{fwt} conserves the number of independent degrees of freedom and is 
linearly invertible. 

In the toy model of $N=2$  we get a mapping 
$
q_0,q_1,q_2,q_3 \to d^1_0, d^1_1, d^2_0, c^2_0,
$ 
with a two-scale decomposition of the initial data:
\begin{align*}
d^1_0 &= \frac{q_0-q_1}{\sqrt{2}},&  d^1_1  &= \frac{q_2-q_3}{\sqrt{2}},\\ 
d^2_0 &= \frac{q_0+q_1-q_2-q_3}{2},& c^2_0 &= \frac{q_0+q_1+q_2+q_3}{2}.
\end{align*}
The last coefficient $c^2_0$ is merely the mean value of the field $q$, which 
can be eliminated in continual limit $N\to \infty$. The wavelet coefficient $d^2_0$ 
is responsible for the global dynamics of the field $q$ on the whole time interval 
$[0,T]$. The coefficients $d^1_0$ and $d^2_1$ are responsible for the dynamics of the 
field $q$ on the left and the right half-intervals, respectively. Thus, instead 
of integration over $q_1$ and $q_2$ (at fixed $q_0,q_3$), we can integrate over $d^1_0,d^1_1$,
keeping fixed the remaining wavelet coefficients. If the coefficients $d^m_n$ are {\em operators}, they can be naturally ordered according to the tree shown in Fig.~\ref{t4:pic}. 

Now we can consider similar hierarchic ordering  in the partition of the square  
$[0,T]\otimes [0,T]$ in the $(x^+,x^-)$ plane, considering $x^+$ as the time and $x^-$ as 
the space variable. In two dimensions to store the information written in each 4 points of 
the $j$ hierarchy level $$(c^j_{2k,2m},c^j_{2k+1,2m},c^j_{2k,2m+1},c^j_{2k+1,2m+1})$$ we need 
4 basic functions:
\begin{equation}
\begin{matrix}
\varphi(x^+)\varphi(x^-) &  \chi(x^+) \varphi(x^-)\cr \varphi(x^+)\chi(x^-) & \chi(x^+)\chi(x^-),
\end{matrix} \label{b2}
\end{equation}
where $\chi(\cdot)$ is the basic wavelet and $\varphi(\cdot)$ is the scaling function. In our case $\chi$ is the Haar wavelet, and $\varphi$ is the indicator function 
of the unit interval, see Fig.~\ref{hp:fig}.

Thus at each next hierarchy level $(j+1)$ we have 4 wavelet coefficients 
\begin{align}\nonumber 
c^{j+1}_{k,m}   &= \frac{c^j_{2k,2m} + c^j_{2k,2m+1}+c^j_{2k+1,2m} +c^j_{2k+1,2m+1}}{2},\\
\nonumber
d^{(1),j+1}_{k,m} &= \frac{c^j_{2k,2m} - c^j_{2k,2m+1}+c^j_{2k+1,2m} -c^j_{2k+1,2m+1}}{2},\\
d^{(2),j+1}_{k,m} &= \frac{c^j_{2k,2m} + c^j_{2k,2m+1}-c^j_{2k+1,2m} -c^j_{2k+1,2m+1}}{2},
\label{h4}   \\
\nonumber d^{(3),j+1}_{k,m} &= \frac{c^j_{2k,2m} - c^j_{2k,2m+1}-c^j_{2k+1,2m} +c^j_{2k+1,2m+1}}{2}.
\end{align}
The first coefficient $c$ is proportional to the mean values of the field in the 4 cells it 
inherits. $d^{(1)}$ is sensitive to the movement in the $x^-$ direction, $d^{(2)}$ is sensitive to 
the movement in $x^+$ direction, and $d^{(3)}$ is sensitive to time-like movement. So, this hierarchic description in terms of the light cone variables $x^+$ and $x^-$ is absolutely 
symmetric with respect to the ''time'' $x^+$ and the ''space'' $x^-$. The movements are allowed in all three directions. The change to the next cell is performed by increasing 
the hierarchy level $j \to j+1$, followed by appropriate localization of the subcell.

To exploit this method we can start with a $(1+1)$-dimensional scalar field theory with $\phi^4$-interaction, written in light-front coordinates, see, e.g.~\cite{CH2020}. Considering the square domain $D=[0,T]\otimes [0,T]$ in the ($x^+,x^-$) plane, and the action functional 
\begin{equation}
S[\phi] = \int_0^T dx^+ \int_0^T dx^- \bigl[
\frac{\partial\phi}{\partial x^+}\frac{\partial\phi}{\partial x^-} 
 - \frac{m^2}{2}\phi^2 - \frac{\lambda}{4!}\phi^4 
\bigr], \label{S2}
\end{equation}
originated from the standard Lagrangian of $\phi^4$ theory,  
We can formally decompose the field $\phi(x^+,x^-)$ into the scale components 
 \begin{equation}
\phi(x^+,x^-) = \sum d_{j,k_1,k_2}^{m_1,m_2} \chi_{j,k_1}^{m_1}(x^+)\chi_{j,k_2}^{m_2}(x^-), 
\end{equation}
where the upper indices $m_1,m_2 \in \{h,g\}$ designate the type of basic function: 
$\chi^h \equiv \varphi, \chi^g \equiv \chi$. Similar decomposition can be written for a full 
four-dimensional case of $\phi(x^+,x^-,\vx_\perp)$. The advantage of the orthogonal wavelet bases 
is that supports of these functions do not have partial intersection: they are either disjoint, or one 
within another. This agrees with the space-time picture of quantum measurements, which can be performed 
either in separate space-time regions, or the state of a subregion is inferred from the measurement 
on the whole region.

Using the bases of orthogonal wavelets in $L^2(\R)$ the summation over all basic functions provides the partition of a unity.
Due to this property, the mass term 
$$
\frac{m^2}{2} \int \phi^2 dx^+ dx^-
$$
turns into a sum of modulus squared wavelet coefficients
$$
\frac{m^2}{2}\sum |d^{m_1 m_2}_{j,k_1 k_2}|^2.
$$
Similarly, for the generative part 
$$
\int J(x)\phi(x)dx^+dx^- \to \sum J_{j,k_1,k_2}^{m_1,m_2} d^{m_1 m_2}_{j,k_1 k_2}.
$$
The other terms in the action \eqref{S2} are expressed in terms of the so-called {\em wavelet 
connection coefficients}. The kinetic term is expressed as a product of two identical 
coefficients in $x^+$ and $x^-$ coordinates:
\begin{align*}
\int \frac{\partial \phi}{\partial x^+} \frac{\partial \phi}{\partial x^-} dx^+ dx^- 
&= -\int \phi \frac{\partial^2}{\partial x^+ \partial x^-} \phi dx^+ dx^- \\
&= -\int d_{j',k_1',k_2'}^{m_1',m_2'} \chi_{j',k_1'}^{m_1'}(x^+)\chi_{j',k_2'}^{m_2'}(x^-)
 d_{j,k_1,k_2}^{m_1,m_2}
 \frac{\partial \chi_{j,k_1}^{m_1}(x^+)}{\partial x^+}
 \frac{\partial \chi_{j,k_2}^{m_2}(x^-)}{\partial x^-} dx^+ dx^- \\
 &= -d_{j',k_1',k_2'}^{m_1',m_2'} d_{j,k_1,k_2}^{m_1,m_2} \Omega_{j',k_1-k_1'}^{m_1',m_1}
 \Omega_{j,k_2-k_2'}^{m_2',m_2}
 \end{align*}
 
These connection coefficients 
$
\Omega_{j,k-k'}^{m',m} := \int dx \chi_{j,k'}^{m'}(x) \frac{\partial \chi_{j,k}^m(x)}{\partial x}
$
are presented in \cite{RL1997,Bulut2016,Polyzou2020}. Thus the whole generating functional 
can be written as the integral over all wavelet coefficients spanning the domain $[0,T]\otimes [0,T]$ in 
the ($x^+,x^-$) plane:
\begin{equation}
Z[J] = \int \cD d^{m_1,m_2,\ldots}_{j,k_1,k_2,\ldots} e^{\frac{\imath}{\hbar}
S[d^{m_1,m_2,\ldots}_{j,k_1,k_2,\ldots}] + \imath d^{m_1,m_2,\ldots}_{j,k_1,k_2,\ldots} J^{m_1,m_2,\ldots}_{j,k_1,k_2,\ldots}} \label{zjc}
\end{equation}
Here the functional integration symbol $\cD$ means the product of differentials of all wavelet coefficients. 
To make the generation functional \eqref{zjc} and corresponding Green functions finite, the set of wavelet 
coefficients should be restricted by certain best resolution scale. The restriction of the number of scales is 
a usual thing in numerical wavelet transform. In case of quantum field theory this is a kind of lattice regularization \cite{Battle1999}.

The wavelet coefficients $d^{m_1,m_2}_{j,k_1,k_2}; m_1,m_2 \in \{h,g\}$ represent 
4 possible types of fields supported by finite-size regions in ($x^+,x^-$) plane. Similar construction for any field $f$ can be presented in terms of 
continuous wavelet transform written in light-front coordinates. Since the 
continuous wavelet transform in Minkowski space is defined separately 
in 4 domains \eqref{pg11}
\begin{align*}
A_1: k^+>0,k^->0 &,& A_2: k^+<0,k^->0&,& A_3: k^+>0,k^->0 &,& A_4: k^+<0,k^-<0, 
\end{align*}
 we have a set of 4 different wavelet coefficients
\begin{equation}
W^i_{ab\eta\phi} = \int_{A_i} e^{\imath k_- b_+ + \imath k_+ b_- -\imath \vk_\perp \vb_\perp} \tilde{f}(k_-,k_+,\vk_\perp) \overline{\tilde{\chi}}(ae^\eta k_-,a e^{-\eta}k_+,aR^{-1}(\phi)\vk_\perp) \frac{dk_+ dk_- d^2\vk_\perp}{(2\pi)^4},
\end{equation}
where $\eta$ is the boost angle, and $R(\phi)$ is rotation matrix \cite{AK2013,AK2013iv}.
However, for technical reasons analytic calculations of loop integrals are easier in Euclidean 
version of the scale-dependent theory described in {\em Section \ref{eucl:sec}}.

\section{Conclusion}
In present paper we have considered the application of wavelet transform to the quantum field theory models written 
in light front variables. Our research have been inspired by a recent paper \cite{Polyzou2020}, where the author used the $x^+$ variable as 'time, for 
time-ordering only, but applied wavelet transform to the remaining 'spatial' coordinates ($x^-,\vx^\perp$) to 
resolve details of different scales. This approach, although being favourable in high-energy physics settings with prescribed beam direction, introduces an asymmetry between $x^+$ and $x^-$, and hence an asymmetry between 
the forward and backward motion along the $x$-axis. 

We have shown that  quantum field theory models can be written in light-front variables in a way totally 
symmetric with respect to $x^+$ and $x^-$. Feynman path integral in our approach turns to be the sum over 
all possible sequences of {\em events} between the initial and the final space-time region supplied 
with operator-valued measure, which describes the quantum field that can be potentially measured on this 
region. The representation of this measure in terms of wavelet transform is symmetric with respect to 
$x^+$ and $x^-$ variables. Its value is limited by the scale parameter -- the best    
resolution of measurement, -- i.e. the minimal size of space-time domain the measurement can be carried on. 
This leads to quantum field theory models finite by construction. 

In our approach the causal ordering of events takes place not only in time ($x^+$ or $x^-$), as usual, 
but also by inclusion: the event $\phi_B$ is constrained by event $\phi_A$ if $B\subset A$, where $A$ and $B$ 
are two space-time regions. As a matter of fact, our approach generalises the notion of space-time event 
happening at $P \in \R^{1,3}$ to more general definition adopted in probability theory.

Starting from the flat Minkowski space and considering a simple toy model
the authors understand that a consistent account of path integration over 
finite-size regions may be given only in terms of quantum gravity theory, where 
different tree-like structures, similar to our construction of discrete wavelet 
transform, are already in use, say in terms of the AdS/CFT correspondence\cite{Watanabe2017}.
This will be the subject of future work. 
\section*{Statements and Declarations}
The authors have no competing interests. All authors have been paid 
from budget - no funding acknowledgement is required.

\section*{Acknowledgement} The authors are thankful to Profs. J.-P. Gazeau, M.Hnatich, S.Mikhailov, M.Perel, and  W.N.Polyzou for useful comments and references.
%
\end{document}